\documentclass[11pt]{iopart}

\usepackage{setspace} 
\usepackage{graphicx}
\usepackage{multicol}

\begin{document}


\title[Contraction of cross-linked actomyosin bundles]{Contraction of cross-linked actomyosin bundles} 

\author{Natsuhiko Yoshinaga $^1$ and Philippe Marcq $^2$}

\address{$^1$  WPI Advanced Institute for Materials Research (WPI-AIMR), 
Tohoku University, Sendai 980-8577, Japan}
\address{$^2$ Laboratoire Physico-Chimie Curie,
Institut Curie, Universit\'e Pierre et Marie Curie, CNRS UMR 168, 75005 Paris, France}
\ead{\mailto{yoshinaga@wpi-aimr.tohoku.ac.jp}, \mailto{philippe.marcq@curie.fr}}
\date{\today}


\begin{abstract}
Cross-linked actomyosin bundles retract when severed \emph{in vivo}
by laser ablation, or when isolated from the cell and 
micromanipulated \emph{in vitro} in the presence of ATP. 
We identify the time scale for contraction as a viscoelastic time 
$\tau$, where the viscosity 
is due to (internal) protein friction.
We obtain an estimate of 
the order of magnitude of the contraction time $\tau \approx 10-100$ s, 
consistent with available experimental data for 
circumferential microfilament bundles and stress fibers. 
Our results are supported by an exactly solvable, hydrodynamic model 
of a retracting bundle as a cylinder of isotropic, active matter,
from which the order of magnitude of the active stress is estimated. 
\end{abstract}


\section{Introduction}
\label{sec:intro}

Cells organise filamentous actin  (F-actin) into complex 
cytoskeletal structures that play a major role in determining cell rheology  
and cell shape, and in generating the forces necessary for motility, cell 
division, or tissue stability \cite{Pollard2007}.
Depending on the nature of the actin-binding proteins that nucleate,
help polymerise, and cross-link actin filaments, the structure of 
F-actin networks varies widely \cite{Stricker2009,Lieleg2010}, 
from the near isotropic gels found in the cell cortex to the linear 
bundles that form filopodia or stereocilia \cite{Furukawa1997,Bartles2000}.
Passive cross-linkers, such as $\alpha$-actinin or fascin,
contribute to the network elasticity, wheras bipolar mini-filaments 
of non-muscle myosin II motors that pull on neighbouring actin filaments 
of opposite polarity are responsible for contractility. 

Cross-linked actomyosin bundles include the circumferential
bundles (CBs) found near apical cell junctions of epithelia
\cite{Cavey2009}, and the stress fibers (SFs) assembled in the cytoskeleton
of non-muscle animal cells that exert substantial traction forces on their
environment \cite{Pellegrin2007}.
They are robust and stable organelles, that can be extracted from the cell 
while maintaining their shape, structure, and contractility,
over durations long enough to allow mechanical testing
\cite{Owaribe1981,Owaribe1982,Katoh1998}.
Indeed isolated circumferential bundles
\cite{Owaribe1982} as well as isolated stress fibers \cite{Katoh1998}
contract \emph{in vitro} in the presence of adenosine triphosphate (ATP).

Laser photoablation stands out as a powerful and versatile tool,
widely used to reveal the distribution, orientation, and 
relative intensity of stress in living cells and tissues
(see \cite{Colombelli2007,Rauzi2011} for recent reviews).
In order to investigate the cytoskeletal basis of the 
mechanical properties of cells and tissues,
the retraction of ablated actomyosin bundles 
has been studied quantitatively by several groups 
\cite{Rauzi2008,Landsberg2009,Fernandez-Gonzalez2009,Kumar2006,Colombelli2009}.
These experiments give unambiguous evidence that the bundles are under tension
\emph{in vivo}, since they retract once severed, and that
the retraction is an active process, since it is blocked by inhibitors of 
contractility \cite{Fernandez-Gonzalez2009,Kumar2006,Russell2009}.
The relaxation dynamics is generally fitted by a single exponential, 
that defines a contraction time, whereas the initial velocity at the
time of ablation gives access to (relative) measurements of
the preexisting bundle tension.
Interestingly, the retraction that follows laser ablation is also relevant in
(unperturbed) physiological conditions since stress fibers 
spontaneously rupture in vivo \cite{Smith2010}.

We use dimensional analysis to estimate the contraction time
of actomyosin bundles, and show that the main source of
dissipation during contraction is protein friction (section \ref{sec:time}).
The mechanical properties of retracting stress fibers
have so far been interpreted with the help of models 
\cite{Colombelli2009,Russell2009,Stachowiak2009,Besser2011}
that emphasise the underlying sarcomeric structure.
One exception is a discrete, numerical model  based 
on the tensegrity hypothesis that accounts for much of the
observed phenomenology \cite{Luo2008}. In section \ref{sec:model},
we introduce and solve a continuous model of a cross-linked actomyosin
bundle, valid on hydrodynamic length scales large compared to the mesh size
of the F-actin network, and where contractility is modeled
by a constant active stress term \cite{Kruse2005,Juelicher2007}.
In section~\ref{sec:active}, comparison with experimental data yields 
the order of magnitude of the active stress 
of stress fibers and circumferential bundles.

\section{Contraction time}
\label{sec:time}

The relaxation of the length of a stress fiber severed \emph{in vivo}
is well fitted by an exponential function of time,
with a contraction time $\tau$ of the order of $1$ to $10$ s,
depending on cell type \cite{Kumar2006,Colombelli2009,Russell2009},
and spatial position within the cell \cite{Tanner2010}. 
Laser ablation of the apical circumferential bundles of epithelial cells 
yields a somewhat longer contraction time of the order of $10$ to $100$ s
\cite{Rauzi2008,Landsberg2009,Fernandez-Gonzalez2009,Farhadifar2007,Cavey2008}.
Retracting bundles were first depicted as Kelvin-Voigt bodies
\cite{Kumar2006}, using a simple phenomenological model consisting in a dashpot 
and a spring in parallel, with a viscoelastic time equal
to the ratio of the viscosity coefficient to the elastic modulus.
However, the origin of the viscosity remains controversial.
Internal viscosity dominates according to \cite{Colombelli2009},
whereas an external drag force is preferred in \cite{Stachowiak2009},
implying a surprisingly large cytosolic viscosity of the order of $10$ Pa s. 
In \cite{Luo2008}, the contraction time closely follows 
the (microscopic) viscoelastic time of components of the fiber,
a parameter of the model.

In section \ref{sec:time:visco}, we estimate 
the orders of magnitude of the viscosity coefficient due to 
internal protein friction, of the elastic modulus of the bundle,
and deduce the order of magnitude of the viscoelastic contraction time due to
protein friction. 
In section \ref{sec:time:poro}, 
using well established numerical values of the cytosolic viscosity, we 
further show that external viscous drag is negligible, and would
lead to a much shorter viscoelastic time, at variance with experiment.
We next evaluate the poroelastic time
that governs the permeation of the cytosol through the F-actin network, 
and discuss its possible relevance to the dynamics of contraction.
Finally, we calculate the order of magnitude of the elastic modulus
and viscosity coefficient of stress fibers and circumferential bundles
(section \ref{sec:time:matprop}).
An order of magnitude is hereafter defined by the integer value, 
obtained by truncation, of the decimal logarithm of the quantity
of interest.

\subsection{Viscoelastic time}
\label{sec:time:visco}

We propose that the main source of dissipation in cross-linked
actomyosin bundles
lies in the binding-unbinding dynamics of cross-linkers on actin filaments.
This dynamics leads to an effective (protein) friction, as
introduced first in the context of rubber friction \cite{Schallamach1963},
and later for actomyosin dynamics in muscle cells \cite{Tawada1991}.

In the linear regime, the protein friction force $\mathbf{F}_p$ 
exerted on one filament is proportional to its
relative velocity $\mathbf{v}$ (with respect to neighbour filaments), 
with a friction coefficient $\zeta_p$:
$\mathbf{F}_p = - \zeta_p \; \mathbf{v}$.
Since the energy dissipated after unbinding was stored
as elastic energy in strained cross-linkers,
an estimate of  the friction coefficient is \cite{Schallamach1963}:
\begin{equation}
  \label{eq:order:zeta}
\zeta_p = n_X \; k_X \;  \tau_X,
\end{equation}
where  $n_X$ is the average number of attached cross-linkers per filament,  
$k_X$ is the spring constant of the cross-linker,
and  $\tau_X$ is the typical time for bond rupture.
We expect that several cross-linker proteins, either passive or active,  would
contribute additively to the friction coefficient:
$\zeta_p = \sum_X \; n_X k_X  \tau_X$.
To our knowledge, the spring constant of $\alpha$-actinin,
arguably the most prevalent passive cross-linker in stress fibers,
has not been measured. For lack of data, only protein friction due to
myosin filaments will be taken into account.
Using the numerical values $n_X \approx 10$,
$k_X  \approx  0.1 \; \mathrm{pN \;nm}^{-1}$ 
for myosin filaments \cite{Neumann1998,Veigel1998},
and equation (\ref{eq:order:zeta}), we obtain the order of magnitude of the
microscopic friction coefficient
$\zeta_p \approx 10^{-3} - 10^{-2} \; \mathrm{N \; s \; m}^{-1}$.

\begin{figure}[t]
\centerline{\includegraphics[width=0.90\textwidth]{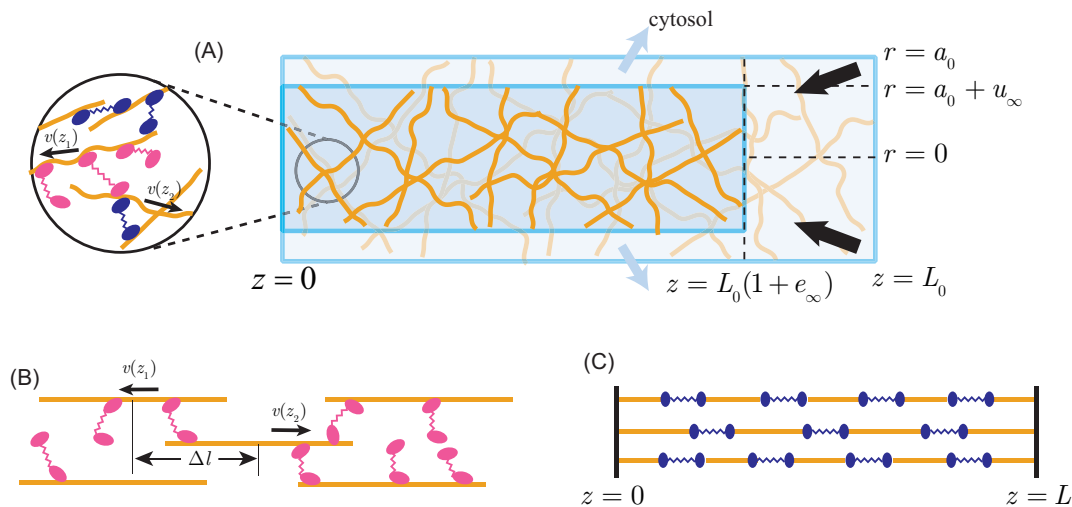}}
\caption{\label{fig:schematics} 
(Color Online) (A) Schematic image of the contraction of an actomyosin bundle.
Permeating cytosol is squeezed out of the network during contraction.
One end of the bundle  is fixed at the origin $z=0$.
The bundle contracts longitudinally from the 
initial length $L_0$ to the final length $L_0 (1 + e_{\infty})$
($e_{\infty} <0$), and radially from the initial radius $a_o$ to the 
final radius $a_0 + u_{\infty}$ ($u_{\infty} <0$).
The velocities of filaments whose center of mass positions are located at 
$z_1$  and $z_2$ are denoted as $v(z_1) = U z_1/L$ and $v(z_2)= U z_2/L$,
respectively. 
Cross-linkers contributing to the bundle viscosity 
(resp. elasticity) are drawn in red (resp. blue).
Importantly, the turn-over of both active motor filaments and passive 
cross-linking proteins gives rise to protein friction.
The typical distance between the centers of mass
of two filaments is $\Delta l \approx l_F$ as schematically shown in (B).
The abstract network structure is sketched in (C) where 
many  pathways connect the two ends ($z=0$ and $z=L$); each pathway
contains filaments and cross-linkers in series.
}
\end{figure}

We expect the dissipative stress $\sigma_p$ due to protein friction 
to scale as $\sigma_p \approx n_F \zeta_p v / A$, 
where $n_F$ is the number of actin filaments in a section
of the bundle, $A$ is the area of the section, and $v$ is the velocity.
In three dimensions, protein friction translates into a viscosity
coefficient $\eta_p$, with  
$\sigma_p = \eta_p \frac{\partial v_z}{\partial z}  
\approx \eta_p \frac{U}{L}$,
where $U$ and $L$ are the contraction velocity and the length
of the bundle. Assuming for simplicity a constant velocity gradient
$v_z = U \frac{z}{L}$, the local relative velocity on the scale of a
filament is $v = U \frac{l_F}{L}$,
where $l_F$ is the typical length of an actin filament 
(see figure~\ref{fig:schematics} (B)).
Note that although we have assumed that filaments are aligned in 
the $z$-direction,
we may also consider an homogeneous, random orientation of filaments. 
This would modify the prefactors which arise from the
average orientation of two filaments. 
Equating the two expressions for $\sigma_p$,
we deduce the order of magnitude of the three-dimensional 
viscosity coefficient:
\begin{equation}
  \label{eq:order:eta}
  \eta_p \approx n_F \; \frac{l_F}{A} \; \zeta_p.
\end{equation}

The stiffness of the bundle is dominated by cross-linkers, such as myosin 
filaments, known to be softer than actin filaments:
$k_X  \approx  0.1 \; \mathrm{pN \;nm}^{-1}$ 
\cite{Neumann1998,Veigel1998} compared to
$k_F  \approx  10 \; \mathrm{pN \;nm}^{-1}$ \cite{Kojima1994,Liu2002}. 
The elasticity of the bundle is characterised by a network structure of
actin filaments and cross-linkers with long residual time. The network
has two ends at $z=0$ and $z=L$ connected by various pathways in
parallel. Each pathway has filaments and cross-linkers connected in
series, and the bundle elasticity is approximated by that of a dominant pathway
(figure~\ref{fig:schematics} (C)). 
When hard and soft springs are connected in series, the
force acting on the ensemble of two springs for the displacement $\Delta
z$ is $F \simeq \Delta z /(k_F^{-1} + k_X^{-1})$ where $k_F$ is a spring
constant of a hard actin filament and $k_X$ is that of a soft
cross-linker. When $k_F \gg k_X$, the force is dominated by $k_X$. 
The elastic stress $\sigma^{\mathrm{el}}$
is of order  $\sigma^{\mathrm{el}}  = E \frac{\partial u_z}{\partial z}
\approx E \frac{\Delta Z}{L}$,
where $E$ and $\Delta Z$ respectively denote the Young modulus, and 
the (macroscopic) displacement of the bundle.
Denoting by $l_X$ 
the typical length between cross-linkers in the network, approximated 
by the length of a myosin filament,
we expect the (microscopic) displacement of a cross-linker to scale as
$\Delta z = \Delta Z \; \frac{l_X}{L}$. Since the number of cross-linkers in a
section is $n_X n_F$, we find for the elastic stress 
$\sigma^{\mathrm{el}} \approx n_X n_F \; k_X \Delta z/A$. Since the
two expressions for $\sigma^{\mathrm{el}}$ are equal, the order
of magnitude of the Young modulus reads:
\begin{equation}
  \label{eq:order:E}
  E \approx n_X n_F \; \frac{l_X}{A} \; k_X.
\end{equation}

Using equations (\ref{eq:order:zeta}-\ref{eq:order:E}), we obtain the
order of magnitude of the viscoelastic time due to protein friction:
\begin{equation}
  \label{eq:order:tau}
  \tau = \frac{\eta_p}{E} \approx \frac{l_F}{l_X} \; \tau_X,
\end{equation}
independent of the extension of the bundle.
Indeed, experiments suggest that the contraction time  is independent of the 
initial width $a_0$ \cite{Kumar2006},
and of the initial length $l_0$ \cite{Colombelli2010}
of the ablated stress fiber, at odds with the prediction
$\tau \propto l_0^2$ of the sarcomere-based model introduced 
in \cite{Stachowiak2009}.

The typical lengths of actin and myosin filaments are respectively
$l_F \approx 1 \; \mu$m 
and $l_X \approx 0.1 \; \mu$m \cite{Langanger1986,Svitkina1989}. 
The association/dissociation rates of myosin light chain
and $\alpha$-actinin have been measured in stress fibers
\emph{in vivo} in fluorescence recovery after photobleaching
assays \cite{Hotulainen2006}. The turn-over times are of order
$\tau_X \approx 1 - 10 \; \mathrm{s}$, consistent with
the unbinding times measured on single actomyosin
bonds under physiological loads \cite{Nishizaka2000,Ferrer2008}.
We obtain for the
order of magnitude of the contraction time $\tau$ the range
\begin{equation}
  \label{eq:val:tau:SF}
  \tau \approx 10^1 - 10^2 \; \mathrm{s},   
\end{equation}
in accord with experimental observations for
contraction \emph{in vivo} after laser ablation, 
as well as \emph{in vitro} after extraction from the cell
\cite{Owaribe1982,Katoh1998}.

\subsection{Poroelastic time}
\label{sec:time:poro}

One obvious alternative source of dissipation is
hydrodynamic friction, due to the viscosity of the cytosol.
At small scale, the longitudinal friction coefficient due to viscous drag 
on an actin filament, modeled as a cylinder of
length $l_F$ and diameter $d_F$, reads 
$\zeta_D = 2 \pi \eta_{\mathrm{cytosol}} l_F/
(\ln (l_F/d_F) + \gamma_{//})$,
where $\eta_{\mathrm{cytosol}}$ is the viscosity of the cytosol,
and  $\gamma_{//}$ is a 
dimensionless number of order $10^{-1}$ \cite{Tirado1979}.
Using $l_F \approx 1 \; \mu$m, $d_F \approx 10$ nm, 
and well-established values for the cytosolic viscosity
$\eta_{\mathrm{cytosol}} \approx  10^{-3} - 10^{-1} \; \mathrm{Pa \; s}$ 
\cite{Charras2009},
we find $\zeta_D \approx 10^{-9} - 10^{-7} \; \mathrm{N \; s \; m}^{-1}$,
negligible when compared to $\zeta_p$. At the scale of the bundle, with 
 $l_B \approx 10 \; \mu$m, $d_B \approx 0.1 \; \mu$m, we obtain
$\zeta_D \approx 10^{-8} - 10^{-6} \; \mathrm{N \; s \; m}^{-1} \ll \zeta_p$.
Viscous drag against the cytosol may thus be neglected.

Further, cytosol is squeezed out of the biopolymer network
during the contraction: permeation also contributes to the dissipation.
The permeation of a solvent through an elastic network has been
studied  in the context of chemical gels
(see  \cite{Doi2009} for a recent review of stress-diffusion coupling),
and has led to interesting insights into, \emph{e.g.}, the mechanics 
of the poroelastic cytoskeleton \cite{Charras2009}.
In a two-component system where fluid cytosol 
(the solvent) permeates an elastic polymer matrix,
the total stress is expressed as the sum 
of the elastic stress of the polymer network,
and of a pressure term for the liquid solvent:  
$\sigma_{ij}^{\mathrm{tot}} = \sigma_{ij}^{\mathrm{el}}   - p \;  \delta_{ij}$,
where $\delta_{ij}$ is the identity tensor 
(see section~\ref{sec:model} for a detailed calculation).

Supplementing this constitutive equation with Darcy's law
for the permeation of the solvent, one obtains
a diffusion equation for the displacement field of the gel,
with a diffusion constant $ D \simeq \kappa E$, 
where $\kappa$ is the permeability coefficient.
Since $\kappa \propto \xi^2/\eta_{\mathrm{cytosol}}$,
where $\xi$ is the mesh size of the network
and $\eta_{\mathrm{cytosol}}$ is the cytosolic viscosity, 
the time scale for relaxation over a typical distance $a_0$,
evaluated as the radius of the bundle, reads:
\begin{equation}
  \label{eq:tau:poro}
  \tau_p =  \frac{a_0^2}{D} 
\approx \left( \frac{a_0}{\xi} \right)^2 \; \frac{\eta_{\mathrm{cytosol}}}{E}.
\end{equation}
From equation (\ref{eq:order:E}), and $A \approx a_0^2$, we expect 
the poroelastic time to be sensitive to the width of the bundle:
$\tau_p \propto a_0^4$. The typical radius of a stress fiber is
$a_0^{\mathrm{SF}} \approx 100$ nm \cite{Deguchi2005}, whereas
the cross-sectional area of a circumferential bundle is of the order
of $A^{\mathrm{CB}} \approx 1 \; \mu\mathrm{m}^2$
\cite{Owaribe1981,Owaribe1982,Rauzi2008,Landsberg2009}, or
$a_0^{\mathrm{CB}} \approx 1 \;\mu$m.
Using $\xi  \approx  10 \; \mathrm{nm}$ \cite{Owaribe1981,Katoh1998}
and $n_F \approx 10$ \cite{Katoh1998}, we obtain:
\begin{eqnarray}
  \label{eq:val:poro:SF}
 \tau_p^{\mathrm{SF}} &\propto&  10^{-6} - 10^{-4} \; \mathrm{s},\\
  \label{eq:val:poro:CB}
 \tau_p^{\mathrm{CB}} &\propto&  10^{-2} - 10^{0} \; \mathrm{s}.
\end{eqnarray}

Permeation of the cytosol through 
the polymer network occurs on diffusive times that are shorter than
the observed contraction time, and may be safely ignored.
It may however become relevant for very thick bundles:
note that the upper bound of the estimate of the poro\-elastic time 
is only one order of magnitude shorter than observed contraction times 
in the case of circumferential bundles.
In section~\ref{sec:model:model}, we introduce and solve a theoretical model of
an actomyosin bundle as an active, viscoporoelastic material.
We show in section~\ref{sec:model:visco} that
the viscoelastic behavior, as discussed in section~\ref{sec:time:visco},  
is recovered in the limit where the viscoelastic time is much larger 
than poroelastic diffusion times.

\subsection{Material properties}
\label{sec:time:matprop}

The Young modulus may be directly estimated using equation (\ref{eq:order:E}).
For stress fibers, $A^{\mathrm{SF}} \approx 10^{-2} \;\mu \mathrm{m}^2$,
whence $E^{\mathrm{SF}} \approx 10^5 \; \mathrm{Pa}$.
In \cite{Deguchi2005,Deguchi2006},
the force-extension curve of an isolated stress fiber has 
been measured in the absence of ATP, up
to large deformations of the order of $200 \%$ where the fiber
breaks. In the linear regime of small strains, 
the fiber's Young modulus was estimated to be of 
the order of $10^5$ to $10^6$ Pa.
Circumferential bundles are thicker 
$A^{\mathrm{CB}} \approx 1 \;\mu \mathrm{m}^2$, and correspondingly softer:
we obtain from equation (\ref{eq:order:E})
$E^{\mathrm{CB}} \approx 10^3 \; \mathrm{Pa}$, somewhat higher than
the value $E^{\mathrm{CB}} \approx 10^2$ Pa given in
\cite{Feneberg2004}, where the elastic modulus of the actin cortex was measured
in vivo in confluent endothelial cells.

Estimates of the bundle viscosity are obtained from equations 
(\ref{eq:order:zeta}) and (\ref{eq:order:eta}). Since: 
\begin{equation}
  \label{eq:order:eta:full}
  \eta_p \approx n_X n_F \; \frac{l_F}{A} \; k_X \; \tau_X,
\end{equation}
we find: 
\begin{eqnarray}
  \label{eq:val:eta:SF}
\eta_p^{\mathrm{SF}} &\approx 10^{6}& - 10^{7} \; \mathrm{Pa \; s},\\
  \label{eq:val:eta:CB}
\eta_p^{\mathrm{CB}} &\approx 10^{4}& - 10^{5} \; \mathrm{Pa \; s}.
\end{eqnarray}
Although the effect of internal friction has been considered
in models of stress fiber retraction \cite{Colombelli2009,Stachowiak2009}, 
this constitutes, to our knowledge, the first calculation of the 
numerical value of the viscosity coefficient.
We emphasise that this value is an order of magnitude, derived 
from dimensional analysis, and defer a more rigorous treatment
to future work.

\section{Models of active contraction}
\label{sec:model}

We propose a three-dimensional, continuous, exactly solvable 
model of a cross-linked actomyosin bundle 
as an active, cylindrical body,
inspired by  active gel descriptions of the cytoskeleton
\cite{Kruse2005,Juelicher2007}, where
chemo-mechanical coupling is taken into account at linear order
by including an  active stress term in the constitutive equations. 
The resulting dynamical equations can be solved analytically, 
and yield results that agree well with experimental data.
Related approaches successfully accounted for the possibility of
oscillations in muscle sarcomeres, described as 
active elastic bodies \cite{Guenther2007,Banerjee2011}
and for the existence of distinct types of polarity patterns 
in contractile actomyosin bundles described as
active, polar elastomers \cite{Yoshinaga2010}. 
The assumptions made are discussed in detail in section \ref{sec:model:intro}.
We introduce in section~\ref{sec:model:model} a theoretical description of
an actomyosin bundle as an active, viscoporoelastic material.
We  discuss in section~\ref{sec:model:poro} the poroelastic limit.
In section~\ref{sec:model:visco},
we show rigorously that the bundle behaves as an active viscoelastic 
body when the viscoelastic time is much larger than the poroelastic 
diffusion times, and solve the model in this limit.

\subsection{General formulation of the model}
\label{sec:model:intro}

We formulate a hydrodynamic description valid on
length scales large compared to the typical mesh size
of the underlying polymer network 
$d \gg \xi  \approx  10 \; \mathrm{nm}$ 
\cite{Owaribe1981,Katoh1998}.
In experiments, bundles are ablated far from their end points.
We consider the retraction
of one half of the severed bundle: the other  
tip is fixed and sets the origin of the $z$ axis.
We consider the simplified geometry where 
(half-) a bundle initially adopts the shape of a cylinder of axis $Oz$
(see figure~\ref{fig:schematics} (A)).
We naturally use the cylindrical coordinates $(r, \theta, z)$.
The bundle initially occupies the volume defined by $0 \le r \le a_0$,
$0 \le z \le l_0$, where $a_0$ and $l_0$ respectively denote 
the initial radius and length. 

While this simplified geometry may seem a natural approximation
for stress fibers spanning the ventral side of adherent cells 
and bound to the substrate at (fixed) focal adhesions, we believe
that the following observations also justify its relevance
for apical circumferential bundles.
First, the ablation is generally performed along nearly straight cell 
junctions: the local curvature may be neglected.
Second, the shape of the cell is most often preserved
despite the ablation of one cell junction, while other 
apical bundles in the same cell remain mostly unperturbed
\cite{Rauzi2008,Fernandez-Gonzalez2009}, perhaps due to the
stabilising influence of the surrounding epithelium. 
The retracting cell junction may therefore be treated as 
attached to an immobile cell-cell vertex and 
nearly independent of the remainder of the apical actin cortex.

The severed stress fibers 
do not retract when the ablation is performed in the presence 
of inhibitors of contractility \cite{Katoh1998,Kumar2006}.
For this reason, the displacement field $\mathbf{u}$ is defined
with respect to the initial state (immediately after ablation):
$\mathbf{u}(t = 0) = \mathbf{0}$.
Following \cite{Yamaue2005}, we postulate the following functional 
form for components of $\mathbf{u}$:
  \begin{eqnarray}
    u_r(r,\theta,z,t) &=& u(r,t)     \label{eq:strain:ur}\\
    u_{\theta}(r,\theta,z,t) &=& 0     \label{eq:strain:ut}\\
    u_z(r,\theta,z,t) &=& e(t) \; z     \label{eq:strain:uz}.
  \end{eqnarray}
This Ansatz is central to our calculation.
The displacement field is independent of $\theta$
due to rotational invariance about the axis $Oz$.
The origin is fixed, $u_z(z = 0) = 0$ at all times,
and the longitudinal strain $e(t)$ is uniform.
The initial cylindrical shape is preserved
by our Ansatz for the displacement field, 
and the radius $a(t)$  and length $L(t)$ 
of the cylinder at a later time $t > 0$ read:
  \begin{eqnarray}
    \label{eq:strain:a}
    a(t) &=& a_0 + u(a_0,t), \\
    \label{eq:strain:L}
    L(t) &=& L_0 \; (1 + e(t)),
  \end{eqnarray}
where $e(t)$ appears as the relative variation of length of the bundle,
negative for a contraction (see figure~\ref{fig:schematics} (A)).

For simplicity, we assume that the bundle material properties are
isotropic. In section \ref{sec:time}, our estimates rely on the physical 
properties of cross-linkers, which remain approximately isotropic.
In addition, the orientation of actin filaments inside actomyosin bundles
shows a significant amount of disorder.
The binding/unbinding kinetics of cross-linkers is taken into account
by an effective, bulk viscous term with coefficient $\eta_p$,
due to protein friction. We include in the constitutive equations 
a constant, isotropic, active stress 
$\sigma_{ij}^{\mathrm{active}} = \sigma_A \delta_{ij}$, 
and $\sigma_A$ is positive since the bundle is under tension.
The active stress $\sigma_A$ is assumed to be uniform, 
and independent of time:
we neglect the banded myosin patterns that reflect the (disordered) 
sarcomeric organisation typical of stress fibers \cite{Peterson2004}.
Note however that not all contractile actomyosin bundles 
exhibit a sarcomeric organisation \cite{Cramer1997,Mseka2009}. 

On the time scale involved, the final rest state is well defined. 
For stress fibers, the relative total retraction length 
$|e_{\infty}|$ is of the order of 
$10 \%$ \cite{Kumar2006,Colombelli2009}:
deformations are small enough for linear elasticity to hold.
We denote by $E$ and $\nu$ the Young modulus and the Poisson ratio 
of the bundle, and by $K = \frac{E}{3 (1 - 2 \nu)}$ and 
$G = \frac{E}{2 (1 + \nu)}$ the bulk and shear elastic moduli.
Shear strains and shear elastic stresses cancel by construction.
The normal elastic stresses read:
  \begin{eqnarray}
 \sigma_{rr}^{\mathrm{el}} &=& 
\frac{E}{(1 + \nu) (1 - 2 \nu)} \; 
\left( (1 - \nu) \frac{\partial u}{\partial r} + 
 \; \nu ( \frac{u}{r} + e ) \right), 
    \label{eq:stress:sigr}\\
 \sigma_{\theta \theta}^{\mathrm{el}} &=& 
\frac{E}{(1 + \nu) (1 - 2 \nu)} \; 
\left( (1 - \nu) \frac{u}{r}  +
 \; \nu ( \frac{\partial u}{\partial r} + e )  \right),
\label{eq:stress:sigt}
 \\
 \sigma_{z z}^{\mathrm{el}} &=& 
\frac{E}{(1 + \nu) (1 - 2 \nu)} \; 
\left( (1 - \nu) \; e + 
 \; \nu ( \frac{u}{r} +  
\frac{\partial u}{\partial r} )     \right).
\label{eq:stress:sigz}
  \end{eqnarray}

Like any organelle in a living cell, actomyosin bundles are highly 
dynamic structures: protein components constantly renew while the
global organisation of the bundle is preserved. 
For stress fibers, fluorescence recovery assays performed \emph{in vivo}
yield an association/dissociation time 
of the order of a few minutes for actin \cite{Hotulainen2006,Campbell2007}.
In this model, we assume that a steady-state is reached over the duration of 
ablation experiments, and accordingly that the
material content of the bundle is constant during relaxation.

\subsection{Active viscoporoelastic contraction}
\label{sec:model:model}

In this section, we study the permeation of a viscous
liquid (the cytosol) through an active, elastic polymer network
with transient cross-links. 
Active permeating gels have recently been studied in
detail \cite{Joanny2007,Callan-Jones2011}.
Here, we follow a simpler approach, inspired by an exact calculation  
pertaining to the free swelling of a chemical gel
whose initial shape is a cylinder \cite{Yamaue2005},
including two additional ingredients: a constant
active stress, and a bulk viscous stress due to protein friction. 

Since the strain field obeys the Ansatz~(\ref{eq:strain:ur}-\ref{eq:strain:uz}),
shear strains and shear elastic stresses cancel. 
The normal stresses read:
  \begin{eqnarray}
 \sigma_{rr} &=& \eta_p \; \frac{\partial v_r}{\partial r}
+ \sigma_A + \sigma^{\mathrm{el}}_{rr},
    \label{eq:stress:poror}\\
 \sigma_{\theta \theta} &=&   \eta_p \; \frac{v_r}{r} +  \sigma_A
+ \sigma^{\mathrm{el}}_{\theta \theta},  
\label{eq:stress:porot} \\
 \sigma_{z z} &=& \eta_p \; \frac{\partial v_z}{\partial z}
+ \sigma_A +   \sigma^{\mathrm{el}}_{z z},
\label{eq:stress:poroz}
  \end{eqnarray}
where $\mathbf{v} = \frac{\partial \mathbf{u}}{\partial t}$ denotes
the network velocity field.
For a constant active stress $\sigma_A$, force balance reads:
  \begin{eqnarray}
\partial_r p  &=&  
 \partial_r \sigma_{rr}^{\mathrm{el}} + 
\frac{1}{r} \left( \sigma_{rr}^{\mathrm{el}} - 
\sigma_{\theta \; \theta}^{\mathrm{el}} \right) +
\eta_p \; \frac{\partial}{\partial r} \left( \frac{1}{r} \; 
\frac{\partial}{\partial r} \left( r v_r  \right) \right), 
    \label{eq:newton:poro:r}\\
\partial_{\theta} p  &=&  \partial_{\theta} \sigma_{\theta \theta}^{\mathrm{el}},
    \label{eq:newton:poro:t}\\
\partial_z p  &=&  \partial_z \sigma_{zz}^{\mathrm{el}},
    \label{eq:newton:poro:z}
  \end{eqnarray}
where $p$ is the pressure field of the cytosol.
Since $\partial_{\theta} \sigma_{\theta \theta}^{\mathrm{el}} = 
\partial_z \sigma_{zz}^{\mathrm{el}} = 0$, we have
$\partial_{\theta} p  = \partial_z p  =0$: the pressure field 
is a function of radius and time only, $p = p(r,t)$. 
The radial dependence of the pressure field is obtained from
equation (\ref{eq:newton:poro:r}) after integration with respect to $r$,
using the boundary condition $p(a,t) = p_{\mathrm{ext}}$, where
$p_{\mathrm{ext}}$ denotes the hydrodynamic pressure in the surrounding cytosol.
We find:
\begin{equation}
  \label{eq:poro:p}
p(r,t) = p_{\mathrm{ext}} +  
\eta_p \;
\left[ 
\frac{1}{r} \; \frac{\partial}{\partial r} \left( r v_r \right)
\right]_{a}^r
+ \frac{(1 - \nu) E}{(1 + \nu) (1 - 2 \nu)} \; 
\left[ 
\frac{1}{r} \; \frac{\partial}{\partial r} \left( r u  \right)
\right]_{a}^r.
\end{equation}
Using the previous equation, we eliminate the pressure field from Darcy's law 
$\vec{\nabla} \cdot \vec{v} = \kappa \; \nabla^2 p$,
where $\kappa$ is the permeability,
and obtain the following equation for the displacement field:
\begin{equation}
  \label{eq:straindiffusion}
\frac{\partial u}{\partial t} + \frac{1}{2} \; 
\frac{\mathrm{d} e}{\mathrm{d}t} \; r = 
D \; \frac{\partial}{\partial r} \left( \frac{1}{r} \; 
\frac{\partial}{\partial r} \left( r u  \right) \right) 
+ l_p^2 \; \frac{\partial}{\partial r} \left( \frac{1}{r} \; 
\frac{\partial}{\partial r} \left( r v_r  \right) \right), 
\end{equation}
where $l_p^2 = \kappa \eta_p$, 
$D = \frac{1 - \nu}{(1 + \nu) (1 - 2 \nu)} \kappa E$ is a
diffusion coefficient, and their ratio 
\begin{equation}
  \label{eq:tauE}
\tau_E = \frac{l_p^2}{D} = 
\frac{(1 + \nu) (1 - 2 \nu)}{1 - \nu} \; \frac{\eta_p}{E}  
\end{equation}
is a viscoelastic time.

The boundary condition at $r = a$, 
$\sigma_{rr}(r = a, t) = - p_{\mathrm{ext}}$, yields:
\begin{equation}
  \label{eq:zeroforcer:poro}
\frac{E}{(1 + \nu) (1 - 2 \nu)} \; 
\left( (1 - \nu) \frac{\partial u}{\partial r}_{|a} +  
 \; \nu \left( \frac{u(a)}{a} + e \right) \right) 
+ \eta_p \; \frac{\partial v_r}{\partial r}_{|a} = - \sigma_A.
\end{equation}
In the case of free contraction, 
the external force applied on the tip of the cylinder
at $z = L$ is equal to
$\int_0^{a}  \mathrm{d}r \; 2 \pi r \; \sigma_{zz}(r,t) = 
- \pi a^2 \; p_{\mathrm{ext}}$.  Combining the two
boundary conditions, we obtain:
\begin{equation}
  \label{eq:poro:forcez:elim}
\tau_G \; \frac{\mathrm{d} }{\mathrm{d} t} 
\left( \frac{u(a)}{a} - e \right)
 = - \left( \frac{u(a)}{a} - e \right),
\end{equation}
where $\tau_G$ is a viscoelastic time based on the
shear elastic modulus: 
\begin{equation}
  \label{eq:tauG}
\tau_G = \frac{\eta_p}{2 G},
\end{equation}
which contributes to the relaxation provided that
$\frac{u(a_0,t=0)}{a_0} \neq e(t = 0)$, 
see also equations (\ref{eq:ode:solr}-\ref{eq:ode:solz}).
Here, the initial strain is equal to zero 
$u(a_0, t = 0) = e(t = 0) = 0$, and the solution of 
equation~(\ref{eq:poro:forcez:elim}) is for all time $t$:
\begin{equation}
  \label{eq:uvse}
  e(t) = \frac{u(a,t)}{a} \simeq \frac{u(a_0,t)}{a_0}.
\end{equation}
The length and the radius of the bundle both
relax in time according to the same functional form since
$e(t) = \frac{L(t) - L_0}{L_0} = \frac{a(t) - a_0}{a_0}$.

Equations (\ref{eq:straindiffusion}-\ref{eq:zeroforcer:poro}-\ref{eq:uvse}) 
are solved by the expansion:
\begin{eqnarray}
  \label{eq:sol:alpha}
e(t) &=& e_{\infty} + \sum_n e_n \; e^{- \; t/\tau_n},\\
  \label{eq:sol:u}
u(r,t) &=&  u_{\infty}(r) + \sum_n u_n(r) \; e^{- \; t/\tau_n},
\end{eqnarray}
where the $\tau_n$ are relaxation times. 
Equation~(\ref{eq:straindiffusion})
yields a differential equation obeyed by the amplitudes $u_n$ and $e_n$:
\begin{equation}
  \label{eq:ode:un}
u_n + \frac{1}{2} e_n r = D \left( \frac{\tau_E - \tau_n}{\tau_n}\right) \; 
\frac{{\mathrm d}}{{\mathrm d} r} \left( \frac{1}{r} \; 
\frac{{\mathrm d}}{{\mathrm d} r} \left( r u_n  \right) \right).
\end{equation}
In the following, we define the lengths $l_n$ by 
$l_n^2 = D \; |\tau_E - \tau_n|$.

In the long time limit, the stationary solution reads:
\begin{eqnarray}
   e_{\infty} &=& - \frac{\sigma_A }{3 K}, 
  \label{eq:alphainfty}\\
   u_{\infty}(r) &=& - \frac{\sigma_A }{3 K} \; r.
  \label{eq:uinfty}
\end{eqnarray}
Remarkably, we predict that actomyosin bundles also 
contract along the radial direction ($a_{\infty}  < a_0$),
independently of the value of the Poisson ratio $\nu$,
since the final radius reads:
\begin{equation}
  \label{eq:a:infty}
a_{\infty}  = a_0 \;\left( 1 + e_{\infty} \right)
= a_0 \;\left( 1 -  \frac{\sigma_A}{3 K} \right).
\end{equation}
Indeed, electron microscopy imaging indicates that isolated stress
fibers contract radially \cite{Katoh1998}.
To our knowledge, this observation has not been confirmed by
optical imaging in a live cell, due to insufficient spatial resolution.

The relaxational dynamics are solved as follows. When $\tau_E > \tau_n$, 
the functions $u_n(r)$ are the solutions of an inhomogeneous, 
modified Bessel equation:
\begin{equation}
  \label{eq:sol:un}
  u_n(r) = b_n I_1\left( \frac{r}{l_n}\right) 
- \frac{1}{2} e_n r,
\end{equation}
where the coefficient $b_n$ is an amplitude, 
and $I_1(x)$ is a modified Bessel function of the first kind.
Since $u_n(a_0) = a_0 e_n$ (see equation (\ref{eq:uvse})),
we obtain $e_n = \frac{ 2 b_n}{3 a_0} \; I_1\left( \frac{a_0}{l_n}\right)$
from equation~(\ref{eq:sol:un}).
We deduce from  equation (\ref{eq:zeroforcer:poro}) that
the relaxation times $\tau_n$ are solutions of the equation:
\begin{equation}
  \label{eq:xn}
\left( \frac{\tau_E}{\tau_n} - 1 \right) \;
\left[
\frac{a_0}{l_n} \; I_1'\left( \frac{a_0}{l_n}\right) -
\frac{1}{3} \; I_1\left( \frac{a_0}{l_n}\right)
\right] =
\frac{4}{3} \; \frac{\nu}{1 - \nu} \; I_1\left( \frac{a_0}{l_n}\right).
\end{equation}

\begin{figure}[t]
\centerline{\includegraphics[width=0.50\textwidth]{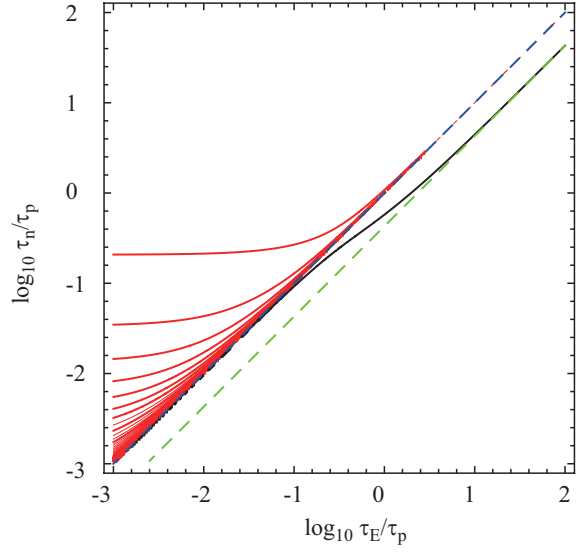}}
\caption{\label{fig:time} 
(Color Online) Dimensionless relaxation times
$\frac{\tau_n}{\tau_p}$ as a function of the parameter
$\frac{\tau_E}{\tau_p}$, obtained numerically for $\nu = 0.4$.
The dashed blue line ($\frac{\tau_n}{\tau_p} = \frac{\tau_E}{\tau_p}$)
separates the two sets of solutions.
The black line gives the unique solution of equation~(\ref{eq:xn}), 
which converges to $\frac{\tau_K}{\tau_p}$ as 
$\frac{\tau_E}{\tau_p} \rightarrow \infty$
(active viscoelastic limit, dashed green line).
The red lines are the solutions of equation~(\ref{eq:poro:xn}), and converge
to constant values as $\frac{\tau_E}{\tau_p} \rightarrow 0$
(active poroelastic limit, solutions of equation~(\ref{eq:xn_porolimit})).
}
\end{figure}

When $\tau_n > \tau_E$, equation~(\ref{eq:ode:un}) is an inhomogeneous 
Bessel equation. Modified Bessel functions are replaced by Bessel functions:
\begin{equation}
  \label{eq:sol:unporo}
  u_n(r) = b_n J_1\left( \frac{r}{l_n}\right) 
- \frac{1}{2} e_n r,
\end{equation}
where $J_1(x)$ is a Bessel function of the first kind,
and $e_n = \frac{ 2 b_n}{3 a_0} \; J_1\left( \frac{a_0}{l_n}\right)$.
The relaxation times $\tau_n$ are solutions of the equation:
\begin{equation}
  \label{eq:poro:xn}
\left( \frac{\tau_E}{\tau_n} - 1 \right) \;
\left[
\frac{a_0}{l_n} \; J_1'\left( \frac{a_0}{l_n}\right) -
\frac{1}{3} \; J_1\left( \frac{a_0}{l_n}\right)
\right] =
\frac{4}{3} \; \frac{\nu}{1 - \nu} \; J_1\left( \frac{a_0}{l_n}\right),
\end{equation}
which may be obtained from equation~(\ref{eq:xn}) 
upon replacing $I_1$ by $J_1$.
In figure~\ref{fig:time}, we plot as a function of the parameter
$\frac{\tau_E}{\tau_p}$ the dimensionless relaxation times
$\frac{\tau_n}{\tau_p}$, at a given value of the Poisson ratio. 
Equation~(\ref{eq:xn}) admits only one solution, whereas 
equation~(\ref{eq:poro:xn}) admits an infinity of 
solutions, which contribute to the expansions (\ref{eq:sol:alpha}-\ref{eq:sol:u}).

\subsection{Active poroelastic limit}
\label{sec:model:poro}

In the limit where $\tau_n \gg \tau_E$, equation~(\ref{eq:poro:xn}) becomes:
\begin{equation}
  \label{eq:xn_porolimit}
\frac{a_0}{r_n} \; J_1'\left( \frac{a_0}{r_n}\right) 
+ \frac{5 \nu- 1}{3(1 - \nu)} \; J_1\left( \frac{a_0}{r_n}\right) =0,
\end{equation}
with $r_n = \sqrt{D \tau_n}$, in agreement with \cite{Yamaue2005}.
The largest solution of equation~(\ref{eq:xn_porolimit}) is 
$\tau_n \approx \tau_p = \frac{a_0^2}{D}$: this limit is realised 
when $\tau_p \gg \tau_E$, as may be possible, \emph{e.g.}, 
for very thick bundles assembled \emph{in vitro},
since we expect the poroelastic time to scale as the fourth
power of bundle radius, see equation~(\ref{eq:tau:poro}) and below.

\subsection{Active viscoelastic limit}
\label{sec:model:visco}

We formally define the viscoelastic limit as the regime where the
permeability coefficient $\kappa$ diverges. Darcy's law
then implies that the cytosolic pressure field is homogeneous
($\mathbf{\nabla} p = \mathbf{0}$).
Since $a_0/l_n \propto 1/\sqrt{\kappa}$, 
we may use $x I'_1(x) \simeq I_1(x)$ and $x J'_1(x) \simeq J_1(x)$ 
 in equations~(\ref{eq:xn}) and (\ref{eq:poro:xn}) 
since $0 < x \ll 1$. For both equations, we obtain
a unique solution  $\tau_n \rightarrow  \tau_K$:
\begin{equation}
  \label{eq:tau}
\tau_K = \frac{\eta_p}{3 K},
\end{equation}
a viscoelastic time based on the bulk modulus $K$.
The expansions (\ref{eq:sol:alpha}-\ref{eq:sol:u}) reduce
to one exponential term. 
Given the initial conditions $u(r, t = 0) = e(t = 0) = 0$,
we find the expression of the displacement fields:
\begin{eqnarray}
  \label{eq:sol:ur}
u_r(r,t) &=& - \frac{\sigma_A }{3 K} \; r \; \left(1 - e^{- \; t/\tau_K} \right),\\
  \label{eq:sol:uz}
u_z(z,t) &=& - \frac{\sigma_A }{3 K} \; z \; \left(1 -  e^{- \; t/\tau_K} \right).
\end{eqnarray}
The length of the ablated bundle 
decreases as an exponential function of time, as observed in experiments.
Equation~(\ref{eq:sol:uz}) also yields
the initial retraction velocity $v_0$, proportional to the active stress:
\begin{equation}
  \label{eq:v0}
v_0 = v_z(z = L_0,t = 0) = -L_0 \; \frac{\sigma_A}{\eta_p}.
\end{equation}

We may rewrite the argument $a_0/l_n$ as a function of time scales: 
$ \frac{a_0}{l_n} = \sqrt{\frac{\tau_p}{|\tau_E - \tau_n|}}$.
We find that $\tau_E - \tau_K = \frac{2 \nu}{1 - \nu}  \tau_K$:
$a_0 \ll l_n$ corresponds to 
$\tau_n \rightarrow \tau_K \approx \tau_E - \tau_K \gg \tau_p$. 
The contraction of an active viscoporoelastic cylinder (with protein
friction) does indeed reduce to an active viscoelastic behavior, 
provided that the viscoelastic time  
is large compared to the poroelastic time.
The cross-over between the two regimes occurs 
when $\tau_p \approx \tau_K$, or $a_0^2 \approx \kappa \eta_p$:
it may be controlled by the bundle radius, with a cross-over radius $a_c$.
Using $\kappa \approx \xi^2/\eta_{\mathrm{cytosol}}$ and
$\eta_p \approx n_F \; \frac{l_F}{a_c^2} \; \zeta_p$ 
(equation (\ref{eq:order:eta})), we find:
\begin{equation}
  \label{eq:ac}
  a_c \approx \left( 
\frac{n_F \, l_F \, \zeta_p \, \xi^2}{\eta_{\mathrm{cytosol}}} 
\right)^{1/4}.
\end{equation}
Using the same numbers as in section \ref{sec:time},
we obtain the value $a_c \approx 10 \; \mu$m, large compared to the 
radius of actomyosin bundles found in cells.
The separation of time scale 
$\tau_K \gg \tau_p$ no longer exists when $a_0 \gg a_c$,
and is absent without protein friction, \emph{i.e.} when cross-linkers
are fixed and do not unbind over the time scale considered.

In \ref{sec:app:visco}, we solve a model of an actomyosin bundle
as an active, viscoelastic material, and check that the results are
identical to the limit presented in this section.
In addition, we consider the case where the initial state differs
from the reference state, and find that the strain,
initially nonzero, relaxes as a linear combination of two exponentials
with characteristic times $\tau_K$ and $\tau_G$, 
equations (\ref{eq:ode:solr}-\ref{eq:ode:solz}).

\section{Active stress}
\label{sec:active}

In experiments, the stationary longitudinal strain is of 
order $|e_{\infty}| \approx 10^{-1}$.
Assuming a value of the Poisson ratio of order
$\nu = 0.4$ \cite{Schmidt1996,Trickey2006}, the estimates of 
section~\ref{sec:time} remain relevant since $3K \approx E$.
Using equation~(\ref{eq:alphainfty}),
we deduce the order of magnitude of the 
active stress $\sigma_A = 3K |e_{\infty}| \approx E |e_{\infty}|$:
\begin{eqnarray}
  \label{eq:val:sigmaA:SF}
\sigma_A^{\mathrm{SF}} &\approx &  10^4 \; \mathrm{Pa},\\
  \label{eq:val:sigmaA:CB}
\sigma_A^{\mathrm{CB}} &\approx &  10^2 \; \mathrm{Pa}.
\end{eqnarray}
Since the stall force of a myosin motor is of order $F_S \approx 1$ pN 
\cite{Howard2005}, the above values agree with a 
naive estimate of the active stress as 
\begin{equation}
  \label{eq:order:sigmaa}
\sigma_A \approx n^A_X n_F \; \frac{F_S}{A},
\end{equation}
where $n^A_X$ is the number of active myosin molecules 
(per filament) that generate stress,
with the same order of magnitude as the total number of attached
cross-linkers per filament $n^A_X \approx n_X$.
Since $L_0 - L_{\infty} = |e_{\infty}| L_0$, 
combining equations (\ref{eq:order:E})  and (\ref{eq:order:sigmaa}) yields: 
\begin{equation}
  \label{eq:deltaL}
  L_0 - L_{\infty}  \approx \frac{F_S}{k_X l_X} \; L_0.
\end{equation}
Experiments suggest that the total retracted length
is independent of the bundle radius $a_0$  \cite{Kumar2006}.

The value of the active stress
obtained for circumferential bundles is close to
the order of magnitude obtained for cortical actin
$ \sigma_A^{\mathrm{CA}} \approx 10^3 \; \mathrm{Pa}$ \cite{Salbreux2009}.
Estimate (\ref{eq:val:sigmaA:SF}) suggests that cells whose stress fibers pull 
on an area of the order of $1 \; \mu \mathrm{m}^2$
exert locally a force of the order  $F_A \approx 10$ nN.
This is consistent with measurements of traction forces exerted by cells 
on micropatterned pillars \cite{Tan2003}.
Adherent cells assemble thicker and more robust stress fibers
when the rigidity of the substrate is large \cite{Ghibaudo2008}.
In this case traction forces saturate to a value independent 
of the external stiffness.
In \cite{Marcq2011}, a simple model of the cytoskeleton as an
active, elastic material allowed to identify active contractility
as the physical origin of the saturation traction force, yielding
$10^4  \; \mathrm{Pa}$ as an order of magnitude of the active stress,
in agreement with (\ref{eq:val:sigmaA:SF}).

The values of $e_{\infty}$ measured in ablation assays
are  similar to the values of the ``preexisting strain''
measured in isolated stress fibers \cite{Deguchi2005}, or of the
``preextension'' upon  unloading the stress fibers of cells adhering 
on a stretchable substrate \cite{Lu2008}.
Indeed, as we have seen in (\ref{eq:alphainfty}-\ref{eq:uinfty}),
actomyosin contraction may be seen as equivalent 
to a reduction in the stress-free reference length of an elastic material
\cite{Deguchi2011}.
Experiments show that the amplitude of the preextension correlates positively 
with contractility \cite{Lu2008}, 
in qualitative agreement with equations (\ref{eq:alphainfty}-\ref{eq:uinfty}).
Further, knock-down of $\alpha$-actinin increases preextension
\cite{Lu2008}. Given that stress fibers depleted in $\alpha$-actinin 
are expected to be softer \cite{Tseng2001}, this observation
also fits with equations (\ref{eq:alphainfty}-\ref{eq:uinfty}),
since a smaller value of the elastic modulus
at fixed active stress leads to a larger strain.

\section{Conclusion}
\label{sec:conc}

Local ablation is a widely used tool to estimate
the value of the local stress in living cells, up to a (generally unknown) 
viscosity coefficient (see equations~(\ref{eq:v0}) and (\ref{eq:v0:elast})).
Ascribing the physical origin of the fiber's viscosity to protein friction, 
itself due to the association/dissociation dynamics of cross-linkers,
we obtain the order of magnitude of the viscosity coefficient $\eta_p$.
Using an independent estimate of the bundle's elastic modulus,
we deduce the order  of magnitude of the contraction time
$\tau \approx 10-100$ s, interpreted as a viscoelastic time.
We find that the order of magnitude of
the poroelastic time scales is much smaller,
and study rigorously the limit where poroelasticity may be neglected
thanks to an exactly solvable model of the contracting bundle.  
We identify a well-defined threshold, controlled for instance by 
the bundle width: viscoelasticity (resp. poroelasticity) dominates when the 
viscoelastic time is much longer (resp. much shorter) than the poroelastic time.
For an isotropic material, the model predicts that contraction
occurs both in the longitudinal and radial directions, irrespective
of the value of the Poisson ratio.
From the observed longitudinal strain,
we deduce the order  of magnitude of the active stress:
$\sigma_A^{\mathrm{CB}} \approx  10^2  \; \mathrm{Pa}$ in circumferential bundles
and $\sigma_A^{\mathrm{SF}} \approx  10^4 \; \mathrm{Pa}$ in stress fibers.
Following the literature, we distinguish stress fibers from circumferential 
bundles. However their protein constituents and macroscopic properties 
are identical: the differences we emphasise may in fact be superficial, 
and turn out to conceal a continuum of parameter values.

We expect our results to be relevant to the mechanics
of other cytoskeletal structures, such as 
the actin purse string at the circumference 
of healing wounds \cite{Tamada2007},
or the supra-cellular actin cable that surrounds the
amnioserosa during dorsal closure of fly embryos  \cite{Rodriguez-Diaz2008}.
Circular, contractile actomyosin bundles also form in adherent (single) 
fibroblasts during the course of spreading \cite{Senju2009}. 
Modeling a retracting bundle as a straight cylinder is relevant
locally along a ring as long as the retracted length is small
compared to the radius. 
It may be interesting to generalise the calculation presented here
from a cylindrical to a toroidal geometry.
Strikingly, ablation experiments performed at the tissue level
in \emph{Drosophila} pupal epithelia, on the scale of one hundred cells, 
reveal a relaxation time of the same order as the contraction
time of individual circumferential bundles \cite{Bonnet2012},
suggesting that the same physical mechanism may be at play.

We strongly believe that estimates of orders of magnitude, based
on dimensional analysis, are a useful contribution to 
quantitative biology \cite{Phillips2009}. 
More quantitative data will be needed in order
to reduce the spread of numerical values.
One promising direction in which our ideas may be tested
is that of bundles reconstituted \emph{in vitro}
\cite{Vignjevic2003,Claessens2006,Strehle2011,Thoresen2011}, 
where microscopic and macroscopic parameters may be varied 
in a controlled fashion. 
For instance, we expect that the contraction time increases 
linearly with the average length of actin filaments $l_F$
(equation~(\ref{eq:order:tau})), a prediction that
may be tested experimentally.
On the theoretical side, one would like to be able to calculate in a rigorous
manner the material  parameters of cross-linked actomyosin bundles 
from the microscopic parameters pertaining to individual constituents
(filaments and cross-linkers).
This important question is left for future study.


\ack
The authors thank Axel Buguin, Damien Cuvelier, Fran\c{c}ois Graner, 
Shinji Deguchi, Jean-Fran\c{c}ois Joanny, Jacques Prost and Tetsuo Yamaguchi 
for fruitful discussions, as well as anonymous referees for 
constructive criticism. 
N.~Y. acknowledges support by a Grant-in-Aid 
for Young Scientists (B) (No.23740317).
P.~M. would like to thank Prof.~Masaki Sano  and Prof.~Takao Ohta
for their kind hospitality, as well as the Kavli Institute for Theoretical 
Physics, with partial support by the National 
Science Foundation under Grant No. NSF PHY05-51164.
This work was supported by JSPS, MAEE and MESR under the Japan-France 
Integrated Action Program (SAKURA).


\appendix

\section{Active viscoelastic contraction}
\label{sec:app:visco}

Using the same Ansatz for the displacement field, 
equations~(\ref{eq:strain:ur}-\ref{eq:strain:uz}), the constitutive equations 
for the active, viscoelastic contraction of a cylinder of isotropic 
material read:
  \begin{eqnarray}
 \sigma_{rr} &=& \eta_p \; \frac{\partial v_r}{\partial r} + \sigma_A + 
\frac{E}{(1 + \nu) (1 - 2 \nu)} \; 
\left( (1 - \nu) \frac{\partial u}{\partial r} + 
 \; \nu ( \frac{u}{r} + e ) \right), 
    \label{eq:stressviscor}\\
 \sigma_{\theta \theta} &=&   \eta_p \; \frac{v_r}{r} +  \sigma_A + 
\frac{E}{(1 + \nu) (1 - 2 \nu)} \; 
\left( (1 - \nu) \frac{u}{r}  +
 \; \nu ( \frac{\partial u}{\partial r} + e )  \right),
\label{eq:stress:viscot} \\
 \sigma_{z z} &=& \eta_p \; \frac{\mathrm{d} e}{\mathrm{d}t} + \sigma_A +
\frac{E}{(1 + \nu) (1 - 2 \nu)} \; 
\left( (1 - \nu) \; e + 
 \; \nu ( \frac{u}{r} +  
\frac{\partial u}{\partial r} )     \right).
\label{eq:stress:viscoz}
  \end{eqnarray}
Again, since  $u$ and $e$ are independent of $\theta$ and $z$, 
the azimuthal and longitudinal components of the force balance
equation are immediately verified. In the radial direction, we find:
\begin{equation}
  \label{eq:visco:Newton}
\eta_p \; \partial_r \left( \frac{1}{r} \partial_r \left( r v_r \right)\right) 
+ \frac{E (1 - \nu)}{(1 + \nu) (1 - 2 \nu)} \;
\partial_r \left( \frac{1}{r} \partial_r \left( r u \right)\right) = 0.  
\end{equation}
A radial displacement linear in $r$:
\begin{eqnarray}
  \label{eq:u}
  u(r,t) = \frac{1}{2} A(t) r
\end{eqnarray}
solves equation~(\ref{eq:visco:Newton}), provided that
the dimensionless function $A(t)$ depends on time only.
The stress field is then uniform (with $\sigma_{\theta \theta} = \sigma_{rr}$): 
  \begin{eqnarray}
 \sigma_{rr}  &=& 
\frac{1}{2} \eta_p \; \frac{\mathrm{d}A}{\mathrm{d}t} + \sigma_A +   
\frac{E}{(1 + \nu) (1 - 2 \nu)} \; 
\left( (1 - \nu) \frac{A}{2} +  \; \nu ( \frac{A}{2} + e ) \right), 
    \label{eq:stress:rt}\\
 \sigma_{z z} &=&  
\eta_p \; \frac{\mathrm{d}e}{\mathrm{d}t} + \sigma_A +   
\frac{E}{(1 + \nu) (1 - 2 \nu)} \; 
\left( (1 - \nu) \; e +  \; \nu A  \right),
\label{eq:stress:z}
  \end{eqnarray}
and determined by the boundary conditions.

Neglecting the external, cytosolic pressure, normal stresses at the free surface are 
$\sigma_{nn \;|r = a} = \sigma_{nn \;|z = L} = 0$. 
We obtain a system of two coupled differential equations for $A(t)$ and $e(t)$:
\begin{eqnarray}
  \label{eq:ode1:r}
(1 + \nu) (1 - 2 \nu) \frac{\eta_p}{E}  \; \frac{\mathrm{d}A}{\mathrm{d}t} &=&
- A - 2 \nu e - 2 (1 + \nu) (1 - 2 \nu) \frac{\sigma_A}{E}, \\
\label{eq:ode1:z}
  (1 + \nu) (1 - 2 \nu) \frac{\eta_p}{E}  \; \frac{\mathrm{d}e}{\mathrm{d}t} &=&
- \nu A + ( \nu -1 ) e + (1 + \nu) (1 - 2 \nu) \frac{\sigma_A}{E},
\end{eqnarray}
diagonalised as:
\begin{eqnarray}
  \label{eq:ode2:r}
\tau_K  \; \frac{\mathrm{d}}{\mathrm{d}t} (A + e) &=&
- (A + e) - \frac{\sigma_A}{K}, \\
\label{eq:ode2:z}
\tau_G  \; \frac{\mathrm{d}}{\mathrm{d}t} (2 e - A ) &=& - (2 e - A),
\end{eqnarray}
where $\tau_G$ and $\tau_K$ are defined in equations (\ref{eq:tauG})
and (\ref{eq:tau}) respectively.
For zero initial displacements, we find $A(t) = 2 e(t)$, and 
$e(t) = - \frac{\sigma_A }{3 K} \; \left(1 -  e^{- \; t/\tau_K} \right)$,
in agreement with equations~(\ref{eq:sol:ur}-\ref{eq:sol:uz}).

If we now take into account a possible difference between the initial state
and the stress-free reference state of the passive bundle, the initial
conditions become $A(t = 0) = A_0 \neq 0$, $e(t = 0) = e_0 \neq 0$. The
strain relaxes as a linear combination of two exponentials with 
characteristic times $\tau_G$ and $\tau_K$:
\begin{eqnarray}
  \label{eq:ode:solr}
A(t) = - \frac{2 \sigma_A }{3 K}  \left(1 -  e^{- \; t/\tau_K} \right)
&+& \frac{2}{3}  \left( A_0 + e_0 \right)  e^{- \; t/\tau_K}
\nonumber\\
&+& \frac{1}{3}  \left( A_0  - 2 e_0 \right)  e^{- \; t/\tau_G},\\
\label{eq:ode:solz}
e(t) = - \frac{\sigma_A }{3 K}  \left(1 -  e^{- \; t/\tau_K} \right)
&+& \frac{1}{3} \; \left( A_0 + e_0  \right) e^{- \; t/\tau_K}
\nonumber \\
&-& \frac{1}{3}  \; \left( A_0  - 2 e_0 \right) e^{- \; t/\tau_G}.
\end{eqnarray}
A nonzero initial elastic stress also contributes to the recoil velocity:
\begin{equation}
  \label{eq:v0:elast}
v_0 = -L_0 \; \frac{\sigma_A + \sigma_{zz}^{\mathrm{el}}(t = 0)}{\eta_p},
\end{equation}
with $ \sigma_{zz}^{\mathrm{el}}(t = 0) = \frac{E}{(1 + \nu) (1 - 2 \nu)} 
\left( (1 - \nu)  e_0 + \nu A_0    \right).$
Since $\frac{\tau_K}{\tau_G} = \frac{1 - 2 \nu}{1 + \nu}$, 
the active contraction of bundles initially stretched or compressed
may yield a measurement of the Poisson ratio.
In \cite{Landsberg2009}, the relaxation of ablated circumferential bundles 
has been fitted by a sum of two exponential functions with relaxation times
$\tau_1$ and $\tau_2$, with $\frac{\tau_1}{\tau_2} \approx 10$.
Assuming that our model is valid, this observation suggests that,
in this experiment, $\frac{\tau_K}{\tau_G} \approx 10$, or equivalently
$\frac{1}{2} - \nu \approx 10^{-1}$.

For completeness, we also solve the case of an incompressible material.
The dynamical equations for $A(t)$ and $e(t)$ read:
\begin{eqnarray}
  \label{eq:ode3:r}
\tau_G  \; \frac{\mathrm{d}A}{\mathrm{d}t} + A  &=&  - \frac{\sigma_A}{G}, \\
\label{eq:ode3:z}
\tau_G  \; \frac{\mathrm{d}e}{\mathrm{d}t} + e &=&  - \frac{\sigma_A}{2 G}.
\end{eqnarray}
We find that the bundle contracts both radially and longitudinally
with a characteristic time $\tau_G$, the viscoelastic time based on 
the shear elastic modulus:
\begin{eqnarray}
  \label{eq:ode3:solr}
A(t) &=& - \frac{\sigma_A }{G} 
+ \left(\frac{\sigma_A }{G}  + A_0 \right) e^{- \; t/\tau_G},\\
\label{eq:ode3:solz}
e(t) &=& - \frac{\sigma_A }{2 G} 
+ \left(\frac{\sigma_A }{2 G}  + e_0 \right) e^{- \; t/\tau_G}.
\end{eqnarray}

\section*{References}
\bibliography{Yoshinaga_contraction}

\end{document}